\documentclass[manuscript,screen]{acmart}

\AtBeginDocument{%
  \providecommand\BibTeX{{%
    \normalfont B\kern-0.5em{\scshape i\kern-0.25em b}\kern-0.8em\TeX}}}

\setcopyright{acmcopyright}
\copyrightyear{2023}
\acmYear{2023}
\acmDOI{XXXXXXX.XXXXXXX}

\acmConference[BIOKDD]{Make sure to enter the correct
	conference title from your rights confirmation emai}{August 09,
	2023}{Long Beach, LA}
\acmPrice{15.00}
\acmISBN{978-1-4503-XXXX-X/18/06}

\usepackage{rotating}
\usepackage{pifont}

\usepackage{booktabs,tabularx,ragged2e}
\usepackage{multirow}
\usepackage{graphicx}
\renewcommand{\arraystretch}{1.4}
\usepackage{soul}
\usepackage{bbm}
\usepackage{algorithm}
\usepackage[noend]{algpseudocode}

\DeclareMathOperator*{\argmax}{arg\,max}

\begin{document}
	
	%%
	%% The "title" command has an optional parameter,
	%% allowing the author to define a "short title" to be used in page headers.
	\title{Redundancy-aware unsupervised rankings for collections of gene sets}
	
	%%
	%% The "author" command and its associated commands are used to define
	%% the authors and their affiliations.
	%% Of note is the shared affiliation of the first two authors, and the
	%% "authornote" and "authornotemark" commands
	%% used to denote shared contribution to the research.
	\author{Chiara Balestra}
	
	\email{chiara.balestra@cs.tu-dortmund.de}
	\affiliation{%
		\institution{TU Dortmund}
		\country{Germany}
	}
	\author{Carlo Maj}
	\affiliation{%
		\institution {Centre for Human Genetics, University of Marburg}
		\country{Germany}
	}
	
	\author{Emmanuel Müller}
	\affiliation{%
		\institution{TU Dortmund}
		\country{Germany}
	}
	
	\email{larst@affiliation.org}
	
	\author{Andreas Mayr}
	\affiliation{%
		\institution{Department of Medical Biometry, Informatics and Epidemiology (IMBIE), University Hospital Bonn}
		\country{Germany}
	}

	%%
	%% By default, the full list of authors will be used in the page
	%% headers. Often, this list is too long, and will overlap
	%% other information printed in the page headers. This command allows
	%% the author to define a more concise list
	%% of authors' names for this purpose.
	\renewcommand{\shortauthors}{Balestra et al.}
	
	%%
	%% The abstract is a short summary of the work to be presented in the
	%% article.
	\begin{abstract}    
		The biological roles of gene sets are used to group them into collections. These collections are often characterized by being high-dimensional, overlapping, and redundant families of sets, thus precluding a straightforward interpretation and study of their content. Bioinformatics looked for solutions to reduce their dimension or increase their intepretability. One possibility lies in aggregating overlapping gene sets to create larger pathways, but the modified biological pathways are hardly biologically justifiable. We propose to use importance scores to rank the pathways in the collections studying the context from a set covering perspective. The proposed Shapley values-based scores consider the distribution of the singletons and the size of the sets in the families; Furthermore, a trick allows us to circumvent the usual exponential complexity of Shapley values' computation. Finally, we address the challenge of including a redundancy awareness in the obtained rankings where, in our case, sets are redundant if they show prominent intersections. \par    
		The rankings can be used to reduce the dimension of collections of gene sets, such that they show lower redundancy and still a high coverage of the genes. We further investigate the impact of our selection on Gene Sets Enrichment Analysis. The proposed method shows a practical utility in bioinformatics to increase the interpretability of the collections of gene sets and a step forward to include redundancy into Shapley values computations. 
	\end{abstract}
	
	%%
	%% The code below is generated by the tool at http://dl.acm.org/ccs.cfm.
	%% Please copy and paste the code instead of the example below.
	%%
	\begin{CCSXML}
		<ccs2012>
		<concept>
		<concept_id>10010520.10010553.10010562</concept_id>
		<concept_desc>Computer systems organization~Embedded systems</concept_desc>
		<concept_significance>500</concept_significance>
		</concept>
		<concept>
		<concept_id>10010520.10010575.10010755</concept_id>
		<concept_desc>Computer systems organization~Redundancy</concept_desc>
		<concept_significance>300</concept_significance>
		</concept>
		<concept>
		<concept_id>10010520.10010553.10010554</concept_id>
		<concept_desc>Computer systems organization~Robotics</concept_desc>
		<concept_significance>100</concept_significance>
		</concept>
		<concept>
		<concept_id>10003033.10003083.10003095</concept_id>
		<concept_desc>Networks~Network reliability</concept_desc>
		<concept_significance>100</concept_significance>
		</concept>
		</ccs2012>
	\end{CCSXML}
	
	\ccsdesc{Unsupervised feature ranking}
	\ccsdesc{Collections of gene sets}
	\ccsdesc{Dimensionality reduction}

	%%
	%% Keywords. The author(s) should pick words that accurately describe
	%% the work being presented. Separate the keywords with commas.
	\keywords{Shapley values, GSEA , feature importance scores, redundancy reduction}

	%%
	%% This command processes the author and affiliation and title
	%% information and builds the first part of the formatted document.
	\maketitle

	\section{Introduction}
	One of the main challenges when working with collections of gene sets is the sheer size, and the following low interpretability of the many gene sets belonging to the same collection. We refer to gene sets or \emph{pathways} as sets of genes deriving from a biological classification of genes concerning chemical or biological functions; these are grouped in variably sized (between hundreds to several thousands of gene sets) collections based on some prior biological or chemical function~\cite{liberzon_molecular_2015}. From this grouping derived scarcely interpretable collections of partly overlapping pathways. \par
	Shapley values~\cite{ref:shapley} allows allocating resources fairly among cooperative game players. In recent years, they found application in feature selection where the 'players', i.e., the features, cooperate in creating a model with high accuracy, and Shapley values help discriminate among relevant and non-relevant features for the label prediction. Shapley values derive their success from the flexible and non-demanding definition of the \emph{value function}, making them an easily applicable tool in various contexts~\cite{rozemberczki2022shapley,lundberg_unified_2017,balestraUSV}. We introduce unsupervised importance scores based on Shapley values for sets within sheer-sized families of sets to reduce their size. Being a function of the distributions of the elements among sets, the scores are (1) positively correlated with the size of the sets and (2) unaware of intersections among them. Hence, we introduce pruning criteria and get new rankings of sets that show no correlation with the sets' sizes and low overlap among sets similarly ranked. As a case study, we apply the method to collections of gene sets: our punished Shapley values affect the correlation among sizes of gene sets and their position in the rankings, although not directly meant to solve this issue. The rankings show excellent behavior regarding the redundancy reduction, and the results suggest that the switch to smaller collections of gene sets does not affect the coverage of the genes. Furthermore, the lower dimensional collections still include a similar number of significant pathways when used for gene set enrichment analysis. As an extension of the paper~\cite{balestra2023redundancy}, we propose an analysis of additional collections of gene sets.

	\section{Ranking sets in families of sets}
	
	% _____________________________________________________________________________
	
	%               COOPERATIVE GAME THEORY
	%               definitions from classical game theory
	%               literature; why it has a broad application
	
	% _____________________________________________________________________________

	A cooperative game is a pair $(\mathcal{N}, v)$ where $\mathcal{N}$ is a finite set of players and $v: 2^{\mathcal{N}} \rightarrow \mathbb{R}_+$ is the so-called \emph{value function}. $v$ maps each subset $ T\subseteq\mathcal{F}$ (usually referred as \emph{coalition}) of players to a real non-negative number $v( T)$ under the hypothesis that $v(\emptyset) = 0$ and $v(\mathcal{N}) = 1$. Shapley values represent one possibility of fairly dividing the payoff of the grand coalition $\mathcal{N}$ among the players such that the amount of resources allocated to each is fair concerning the player's contribution in any possible coalition within the game. The Shapley value of $i$ is the average of its \emph{marginal contributions} $v( T\cup\{i\})-v( T)$ across the possible coalitions $ T\subseteq\mathcal{N}$, i.e.,
	\begin{equation*}
		\phi_i(v)= \sum_{ T\subseteq \mathcal{N}\setminus\{i\}} \frac{(N-t-1)!\ t!} {N!}\cdot (v( T\cup\{i\})-v( T))
	\end{equation*} 
	where $N = |\mathcal{N}|$ and $t = | T|$. The formula requires computing $2^N$ times the value function; instead, microarray games, as a special case of \emph{Sum-Of-Unanimity Games} (SOUG), admit a polynomial closed-form solution.

	% _____________________________________________________________________________ 
	
	%               SOUG GAMES
	
	%              why do we need them? bc they are the base on which 
	%        the computation of microarray games is constructed. 
	
	%        in particular, we get the matrix B, then construct the
	%        corresponding SOUG game v* and then we compute the SV
	
	% _____________________________________________________________________________ 
	
	% _____________________________________________________________________________ 
	
	%    C O M P A R I S O N   W I T H  B O N A S S I
	
	% \textcolor{red}{In Bonassi et al. 2007, they compute the importance of genes given the samples in which the genes are abnormal. So, basically we have gene $g$, all samples, for each sample check if it abnormal in the gene $g$ and put zeroes and ones accordingly. They get an importance of genes. Instead we do: we have a pathway $P$, all genes in the gene set, for each gene check if it is present in the pathway $P$ and put zeroes and ones accordingly. Thus we get an importance of pathways.}
	
	% _____________________________________________________________________________ 
	
	\subsection{The computation of Shapley values for microarray games} 
	
	Consider $\mathcal{F}= \{P_1, \ldots, P_N\}$ a family of sets. We denote with $G= \bigcup_{i\in\{1,\ldots, N\}}P_i$ the elements belonging to at least one set $P_i$ and $M= |G|$. Starting from $\mathcal{F}$ and $G$, we build a binary matrix $B\in \{0,1\}^{N\times M}$ where $B_{ij} = 1$ if $g_j \in P_i$ and $B_{ij} = 0$ otherwise. Transposing the definition given by Moretti et al.~\cite{moretti_class_2007}, for each element $g_j\in G$, we look at the set of sets in which $g_j$ is present, and the information on $g_j$ is conveyed by the column $B_j$ of $B$. We define the \emph{support of $B_j$} as the set $\text{sp}(B_j) = \{P_i \in \mathcal{F} \mid  g_{j}\in P_i\}$. The \emph{microarray game} is then defined as the cooperative game $(\mathcal{F}, v^*)$ where 
	\begin{equation}\label{eq:shapley_microarray}
		v^*( T) = \frac{|\{g_j\in G\mid \text{sp}(B_j)\subseteq  T \text{ and } \text{sp}(B_j)\neq\emptyset\}|}{|G|}.
	\end{equation}
	Hence,  $v^*$ is the ratio of the number of genes' supports that $T$ contains and the number of elements in $G$. As $|G|$ is fixed, $v^*$ is proportional to the number of supports $T$ contains; higher scores are achieved by sets covering the full distribution among sets of a high number of elements. Following Sun et al.~\cite{sun_game_2020} approach, the value function can be easily expressed as a SOUG game such that the computation of Shapley values has polynomial runtime. 
	
	Shapley values assign similar scores to redundant players; hence, similar players are ranked in close positions. We integrate into Shapley values a redundancy-awareness concept such that the player ranked at the $(i+1)$-th place is the least overlapping possible with the first $i$-ranked players $\{P_1,\ldots, P_i\}$ introducing pruning criteria for players highly correlated to the previously ranked ones. 
	\begin{figure*}[!t]
		\centering
		\includegraphics[width=\linewidth]{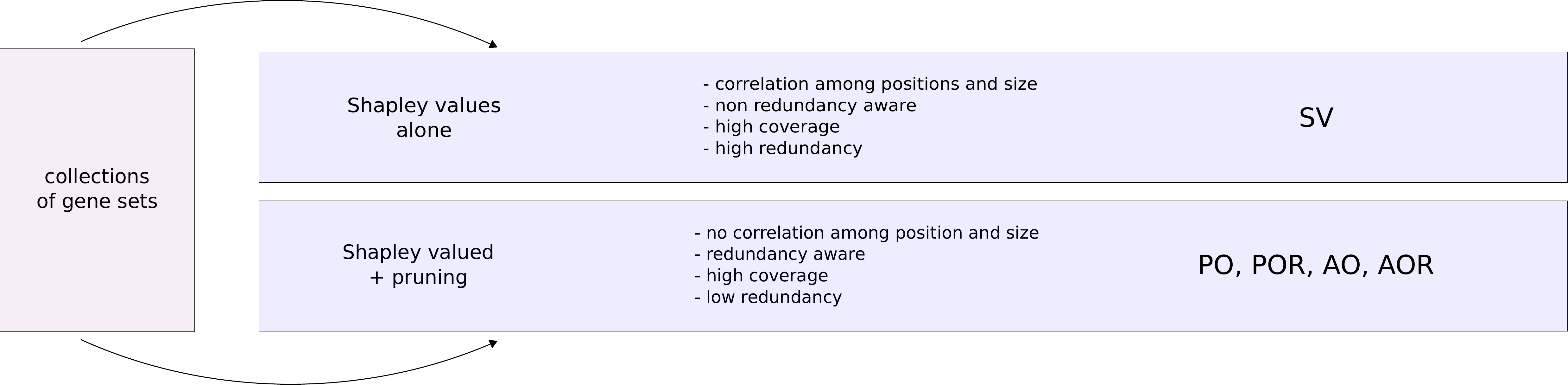}    
		\vspace*{-8mm}
		\caption{\label{fig:genealogical_graph}Shapley values based rankings and properties; the abbreviations refer to Section~\ref{sec:rankings_different}.}
		\vspace{-2mm}
	\end{figure*}
	The sets in $\mathcal{F}$ are arbitrarily large and can overlap, and each $P_i$ contains a variable number $M_i$ of elements. We construct a microarray game based on the binary matrix $B \in \{0,1\}^{N\times M}$; Each row of $B$ represents a set $P_i$, and each column $i$ represents the partial ordering relationship of the element $g_j$ belonging to the sets. The Shapley value of $P_i\in\mathcal{F}$ is computed following two steps~\cite{sun_game_2020}: (1) from the matrix $B$, we get $\mathcal{A} = \{\text{sp}(B_j)\}_{j\in M}\subseteq\mathcal{P}(\mathcal{F})$; (2) then each Shapley value is computed through the formula: 
	\begin{equation}\label{eq:shapley_definition}
		\phi(P_i) = \frac{1}{M}\cdot\sum_{j = 1}^M\left(\mathbbm{1}(P_i \in\, \text{sp}(B_j))\cdot\frac{1}{|\text{sp}(B_j)|}\right).
	\end{equation}
	We use the computed Shapley values as importance scores to rank the $P_i$s: The higher the score of $P_i\in\mathcal{F}$, the more important the set is in the microarray game defined. The importance scores quantify the number of elements $g$ contained in $P_i$, which are also often contained in other sets. $\phi(P_i)$ is a real number in $[0,1]$ and we know that $\sum_{j=1}^N \phi(P_j)=1$ from Shapley values' properties. As mentioned, Shapley values alone are unaware of possible \emph{overlaps} among players.

	% _____________________________________________________________________________ 
	% CORRELATION
	
	\subsection{Low redundancy and high coverage}\label{sec:jaccard_definition}
	We aim for a redundancy-aware ranking of sets in families of sets where two sets are \emph{redundant} if they contain many shared elements. We use as redundancy measure the \emph{Jaccard index}: Given two sets $A$ and $B$, their Jaccard index is $j(A,B) = \frac{|A\cap B|}{|A \cup B|}\in [0,1]$. \par
	While reducing the redundancy among sets, we want to maintain the \emph{coverage} of $G$. We hereby define various pruning criteria and compare them w.r.t. coverage and redundancy of the obtained family of set $S$, using as metrics the \emph{generalized Jaccard index} and the \emph{coverage}. \par
	\textbf{Generalized Jaccard index} -- it is defined as the average Jaccard index among pair of sets in $ S$ and measures the redundancy in $S$:
	\begin{equation}\label{eq:jaccard_rate}
		\text{jac}( S) = \frac{1}{| S|(| S|-1)}\sum_{P_i,P_j \in S, P_i \neq P_j} j(P_i, P_j).
	\end{equation} \par
	\textbf{Coverage of $G$} --  it is the percentage of elements $g\in G$ that are included at least in one set in $ S$, i.e., 
	\begin{equation}\label{eq:coverage}
		c_{G}( S) = {|\cup_{P_i\in S}P_i|\cdot}\frac{100}{|G|}
	\end{equation}
	
	The ranking given by Shapley values tends to rank bigger sets first. Moreover, bigger sets are more likely to overlap. The punishment criteria will also affect the association between the position in the rankings and the size of the sets.

	%%%%%%%%%%%%%%%%%%%%%%%%%%%%%%%%%%%%%%%%%%%%%%%%%%%%%%%%%%%%%%%%%%%%%%%%%%%%%%%%%%
	
	%                   D I F F E R E N T         P U N I S H M E N T S
	
	%%%%%%%%%%%%%%%%%%%%%%%%%%%%%%%%%%%%%%%%%%%%%%%%%%%%%%%%%%%%%%%%%%%%%%%%%%%%%%%%%%

	\subsection{Different punishments criteria}\label{sec:rankings_different}

	We introduce various pruning criteria and provide a detailed comparison of the rankings obtained using collections of gene sets as particular cases (Section~\ref{sec:case_study}). We obtain five different rankings of the sets through a greedy ranking process. 
	
	% _____________________________________________________________________________ 
	
	%                                      S V
	
	\textbf{Shapley values SV}, the original ranking given by Equation~\eqref{eq:shapley_definition} that favors larger sets, and it is unaware of overlaps among sets. Because of the tendency of ranking first larger sets, the ranking obtains high coverage of $G$ when selecting a small number of sets.
	
	% _____________________________________________________________________________ 
	
	%                                      P O
	\textbf{Punished Ordering PO} -- given $S_1(P) =\phi_1(P)$, at each step $n$, the score $S_{n+1}(P)$ obtained by a not yet ranked set $P$ is given by the following recursive formula
	\begin{equation*}
		S_{n+1}(P)=\phi_{n+1}(P) -  \sum_{i=1}^n j(\arg\max_{\bar{P}} S_i(\bar{P}), P), \text{if } n\geq 0;
	\end{equation*}
	At the step $n+1$, the algorithm ranks the set $\tilde{P}_{n+1}=\arg\max_PS_{n+1}(P)$. The Shapley values are re-computed after each iteration to satisfy the efficiency property where $\phi_n(\cdot)$ is the Shapley value at the $n$th iteration.
	
	% _____________________________________________________________________________ 
	
	%                                      P O R
	\textbf{Punished Ordering with Rescaling POR} -- at each iteration, POR rescales the punishment of PO, i.e., $ \sum_{i=1}^n j(\arg\max_{\bar{P}} S_i(\bar{P}), P)$, to the interval $[0, \max_P\{\phi_i(P)\}]$. 
	
	% _____________________________________________________________________________ 
	
	%                                      A O
	
	\textbf{Artificial Ordering AO} -- we introduce an artificial set $AP_n$ updated at each iteration representing the already-ranked sets' union, i.e., $AP_n = \cup_{i=0}^{n}\argmax_{P} S_i(P)$ at the $n$th iteration. The importance score is defined by uniquely pruning with the Jaccard index among the new set $P$ and $AP_n$. In AO, the punishment depends on the elements in $G$ already covered by the first $n$ ranked sets.  Introducing the artificial pathway $AP$ containing the already covered elements avoids multiple punishments for the same overlapping elements typical of PO and POR.
	
	% _____________________________________________________________________________ 
	
	%                                    A O R
	
	\textbf{Artificial Ordering with Rescaling AOR} -- AOR rescales the punishment of AO, i.e., $j(AP_n, P)$ to the interval $[0, \max_P\{\phi_i(P)\}]$.
	
	Note that each criterion orders first the set with the highest Shapley value. The orderings differ from each other from the second-ranked set. The choice of which particular punishment to use is highly dependent on the goal, and there is not a \textit{unique correct} way of choosing which ranking to use. Table~\ref{fig:genealogical_graph} summarizes the properties of the different rankings.%In the Appendix, the pseudo-code to construct the various orderings is provided and the implementation code is available at \texttt{https://github.com/chiarabales/genesetSV}. 

	%%%%%%%%%%%%%%%%%%%%%%%%%%%%%%%%%%%%%%%%%%%%%%%%%%%%%%%%%%%%%%%%%%%%%%%%%%%%%%%%%%
	
	%                               C A S E    S T U D Y
	
	%%%%%%%%%%%%%%%%%%%%%%%%%%%%%%%%%%%%%%%%%%%%%%%%%%%%%%%%%%%%%%%%%%%%%%%%%%%%%%%%%%

	\section{Application to collections of gene sets}\label{sec:case_study}
	
	\begin{table*}[!t]
		\small
		\centering
		\newcolumntype{C}{>{\centering\arraybackslash}X}
		\newcolumntype{R}{>1.5{\centering\arraybackslash}X}
		\begin{tabularx}{\textwidth}{@{}p{0.8cm}CCCCCCCC@{}}
			&& {\textbf{SV}} & {\textbf{PO}} & {\textbf{POR}} & {\textbf{AO}} & {\textbf{AOR}}\\
			\midrule
			\textbf{KEGG13} && 0.691 &-0.251 &0.336 &0.295 &0.344 \\
			\textbf{BIOCARTA}&& 0.32 &-0.372 &0.058 &-0.17 &0.02 \\
			
			\textbf{COVID}&& 0.713 &-0.352 &0.41 &-0.616 &0.401 \\
			
			\bottomrule
		\end{tabularx}
		\caption{\label{tab:correlation} \textbf{Correlation among size and position in the rankings.} Kendall's $\tau$ coefficients to measure the correlation among the position in the ranking and the size of the gene sets.}
		\vspace{-8mm}
	\end{table*}
	
	We apply our methods to gene sets, i.e., lists of genes based on a potential common biological functionality. Typically, the gene sets are grouped in collections of gene sets according to some prior biological knowledge, e.g., involvement in the same biological processes, presence of common biochemical mechanisms, or shared associations with a phenotype~\cite{liberzon_molecular_2015}. We selected collections of gene sets, i.e., KEGG13, BIOCARTA, and COVID, from the \href{https://maayanlab.cloud/Enrichr/#libraries}{Enrich library} and some \href{http://twas-hub.org/traits/}{clinical traits} to assess whether associations exist between gene sets and traits. Each selected collection of gene sets is a set of pathways whose size ranges between $200$ and $250$; the association traits considered contain at least $500$ relevant genes. The implementation code is available on~\href{https://github.com/chiarabales/geneset_SV}{Github}. The results here presented summarize both the three collections of genes sets (KEGG13, BIOCARTA, and COVID) and the ones discussed in Balestra et al.~\cite{balestra2023redundancy}\par
	\begin{table*}[!t]
		\centering
		\small
		\centering
		\newcolumntype{C}{>{\centering\arraybackslash}X}
		\begin{tabularx}{\textwidth}{@{}p{0.8cm}CCCCCCCCCCCc@{}}
			& \multicolumn{3}{c}{\textbf{KEGG13}} & \multicolumn{3}{c}{\textbf{BIOCARTA}} & \multicolumn{3}{c}{\textbf{COVID}} 
			\\ \cmidrule(lr){2-4} \cmidrule(lr){5-7}\cmidrule(lr){8-10}
			\% & 10 & 20 & 40 & 10 & 20 & 40 & 10 & 20 & 40 \\
			\midrule
			\textbf{SV}   & \underline{0.00} & \underline{0.00} & \underline{0.08} & 1.0 & 1.71 & 1.57    & 1.58 & 1.76 & 2.14\\
			\textbf{PO}   & 0.50 & 0.54 & 0.36  & \underline{0.00} & \underline{0.06} & \underline{0.41}    & \underline{0.01} & \underline{0.10} & \underline{0.37}\\
			
			\textbf{POR}   & 0.58 & 0.70 & 0.58  & 0.64 & 0.49 & 0.79
			& 1.01 & 1.15 & 1.18\\
			
			\textbf{AO}   & 0.94 & 0.63 & 0.46  & 1.06 & 1.03 & 0.80    & 0.42 & 1.54 & 1.12\\
			
			\textbf{AOR}  & 0.69 & 0.88 & 0.88  & 0.8 & 0.97 & 1.03    & 1.29 & 1.13 & 1.10 \\
			
			\midrule
			\textbf{SV}  & \underline{54.46} & 73.02 & \underline{89.38} & \underline{41.80} & \underline{62.58} & \underline{83.74} & \underline{46.98} & \underline{70.50} & \underline{88.35}\\
			
			\textbf{PO}  & 43.17 & 48.96 & 69.66 & 31.25 & 48.78 & 72.68 & 11.47 & 23.30 & 63.6\\
			
			\textbf{POR}  & 54.36 & \underline{73.92} & 86.89 & {39.20} & 61.99 & 81.22 & 46.75 & 68.64 & 88.07\\
			
			\textbf{AO} & 31.48 & 42.27 & 72.88 & 27.84 & 42.54 & 64.29 & 13.29 & 15.76 & 34.09\\
			
			\textbf{AOR}  & 46.97 & 59.16 & 78.31 & 35.78 & 53.38 & 75.80 & 42.35 & 63.26 & 84.00\\
			
			\bottomrule
		\end{tabularx}
		\caption{\textbf{Redundancy reduction.} Rescaled generalized Jaccard index of the first $10$, $20$, and $40\%$ of the gene sets; underlined text indicates the minimum generalized Jaccard indices in each column. \textbf{Cumulative coverage of gene sets.} Coverage of the genes when selecting respectively the $10$, $20$ and $40\%$ of the pathways. In each column, underlined text highlights the highest coverage.}
		\vspace{-10mm}
		\label{fig:jaccard_rate}
	\end{table*}
	Our rankings represent a new framework to reduce the dimension of collections of gene sets while keeping a high coverage of genes. As previously stated, collections of gene sets of biologically relevant sets of pathways and pathways $P_i$s are sets of genes $g_j$s; therefore, our approach directly applies to this context. We present the results for collections of gene sets and some phenotypic traits. We obtain \par
	\textbf{no favoritism towards larger pathways:} The Shapley value function assigns to each set $P$ a positive real number incorporating information on the distribution of its elements in the other sets of $\mathcal{F}$. This leads to a positive correlation with the sizes of pathways, i.e., sets with higher dimensions are more likely to get a higher Shapley value. Using Kendall's $\tau$ scores to measure the ordinal association between the pathways' size and the position in the rankings, we show that using AO and PO, this correlation effect is reversed in most of the collections of gene sets; In AOR and POR, there is no clear tendency of a correlation among the dimension of pathways and the position in the ranking (see Balestra et al.~\cite{balestra2023redundancy} and Table~\ref{tab:correlation}). The punishments indirectly affect the strength of the positive correlation between ranking position and their size, as higher-dimensional pathways are more likely to show overlaps among them. \par
	\textbf{rankings of the original pathways without modifying them:} In contrast to usual dimensional reduction methods for collections of gene sets, the pathways are selected and not modified, thus preserving their biological meaning. \par
	\renewcommand{\arraystretch}{0.95}
	\setlength{\tabcolsep}{4.5pt}
	
	\begin{table*}[t]
		\begin{tabularx}{1.05   \linewidth}{cc|cccccccccccccccccccccccccccccccccccccc}
			& & \rotatebox[origin=c]{90}{ HEIGHT } & \rotatebox[origin=c]{90}{ B. PLATELET C. } & \rotatebox[origin=c]{90}{ STANDING H. } & \rotatebox[origin=c]{90}{ B. RED C. } & \rotatebox[origin=c]{90}{ HEEL TSCORE } & \rotatebox[origin=c]{90}{B. WHITE C. } & \rotatebox[origin=c]{90}{ B. EOSINOPHIL C. } & \rotatebox[origin=c]{90}{ SITTING H. } & \rotatebox[origin=c]{90}{ TRUNK PRED. M. } & \rotatebox[origin=c]{90}{ W. BODY FAT FREE M. } & \rotatebox[origin=c]{90}{ W. BODY WATER M. } & \rotatebox[origin=c]{90}{ BMR } & \rotatebox[origin=c]{90}{ IMP. OF W. BODY } & \rotatebox[origin=c]{90}{ BMI } & \rotatebox[origin=c]{90}{ COMP. H. SIZE AGE 10 } & \rotatebox[origin=c]{90}{ ARM PRED. M. R } & \rotatebox[origin=c]{90}{ ARM FAT FREE M. R } & \rotatebox[origin=c]{90}{ LEG FAT FREE M. R } & \rotatebox[origin=c]{90}{ LEG PRED. M. R } & \rotatebox[origin=c]{90}{ IMP. OF LEG R } & \rotatebox[origin=c]{90}{ LUNG FVC } & \rotatebox[origin=c]{90}{ WHRATIO } & \rotatebox[origin=c]{90}{ HAIR PIGMENT }  \\
			
			\midrule
			\multirow{6}{*}{\rotatebox[origin=c]{90}{KEGG13}} & SV &1 & 1 & - & 2 & - & - & - & - & 1 & - & - & - & - & - & - & - & - & - & - & - & - & - & -  \\
			& PO & 2 & - & 2 & 1 & - & 1 & - & - & 6 & - & - & - & - & - & 1 & - & - & - & - & - & - & - & -  \\
			& POR & 1 & - & 1 & 1 & 3 & - & - & - & 1 & - & - & - & - & - & - & - & - & - & - & - & - & - & -  \\
			& AO & 2 & 1 & 3 & 1 & 5 & 1 & - & - & 6 & - & - & - & - & - & 1 & - & - & - & - & - & - & - & -  \\
			& AOR & 1 & - & 1 & 1 & 1 & - & - & - & 1 & - & - & - & - & - & - & - & - & - & - & - & - & - & -  \\
			&ALL& 7& 1& 3& 1& 3& 1& - & - & 8& - & - & - & - & - & - & - & - & - & - & - & - & - & - \\
			
			\midrule
			\multirow{6}{*}{\rotatebox[origin=c]{90}{BIOCARTA}} &SV& - & - & - & - & - & - & - & - & - & - & - & - & - & - & - & - & - & - & - & - & - & - & -  \\
			& PO & 2 & - & 2 & 2 & 1 & - & - & - & 1 & - & - & 1 & - & - & - & - & - & - & - & - & - & - & -  \\
			& POR & 1 & - & 1 & 2 & 1 & - & - & - & - & - & - & 1 & - & - & - & - & - & - & - & - & - & - & -  \\
			& AO & 1 & - & 1 & 1 & 1 & - & 2 & - & 1 & - & - & - & - & - & - & - & - & - & - & - & - & - & -  \\
			& AOR & 2 & - & 2 & 4 & 1 & - & - & - & 1 & - & - & 1 & - & - & - & - & - & - & - & - & - & - & -  \\
			&ALL & 2& - & 2& 2& 1& - & - & - & 1& - & - & - & - & - & - & - & - & - & - & - & - & - & - \\
			
			\midrule
			\multirow{6}{*}{\rotatebox[origin=c]{90}{COVID}} & SV & 9 & 5 & 4 & 5 & 1 & 5 & 6 & - & - & - & - & - & - & - & 4 & 2 & 2 & - & - & - & - & - & 4  \\ 
			& PO & 22 & 15 & 5 & 26 & 4 & 6 & 11 & - & 6 & 2 & 3 & 1 & - & 2 & 3 & 4 & 4 & - & - & - & 2 & 2 & 1  \\ 
			& POR & 12 & 7 & 7 & 9 & 2 & 3 & 4 & - & 6 & 6 & 5 & 6 & 4 & - & 1 & 3 & 2 & 3 & 3 & 3 & 5 & 2 & 3  \\ 
			& AO & 14 & 6 & 6 & 7 & - & 4 & 5 & 1 & 6 & - & - & - & - & 2 & - & - & - & - & - & - & - & - & -  \\
			& AOR & 18 & 8 & 5 & 8 & 3 & 6 & 1 & 1 & 3 & 4 & 3 & 4 & 5 & - & 4 & 1 & 2 & 3 & 3 & 4 & - & 2 & 5  \\
			& ALL & 64& 32& 27& 44& 14& 13& 16& 5& 16& 12& 10& 11& 5& 4& 10& 3& 5& 6& 6& - & 2& 5& 9\\
			\bottomrule
		\end{tabularx}
		\vspace*{0mm}
		\caption{\textbf{Comparison of the number of significant pathways.} Number of significant pathways found using Fisher Exact Test and FDR correction when using the $40\%$ of the pathways in the collection of gene sets ranked using the proposed orderings (namely, SV, PO, POR, AO, and AOR) and the complete collection of gene sets (ALL).} \label{tab:enrich_comparison}
		\vspace{-8mm}
	\end{table*}
	\textbf{evident redundancy reduction:} The pathways' selections deriving from the rankings proposed show much lower overlap than the selection given by the Shapley values alone. We evaluate the redundancy reduction using the generalized Jaccard index. We rescale the generalized Jaccard indexes to the \emph{maximum generalized Jaccard index}, i.e., the maximum Jaccard index among any pairs of pathways within the collection of gene set. The lower the generalized Jaccard index, the better the ranking performs in selecting non-overlapping pathways. PO and AO use strong punishments such that highly overlapping pathways with the first are ranked far from each other. KEGG13 represents an exceptional case where SV achieves low redundancy without additional pruning criteria. Table~\ref{fig:jaccard_rate} shows a direct comparison for the $10$, $20$ and $40\%$ of the pathways. \par
	\textbf{high coverage of the genes:} We investigate the coverage of the genes {in percentage} when only considering the first $n$ ranked pathways using the various rankings. SV, POR, and AOR outperform the rankings PO and AO give. The high coverage achieved by SV is due to the correlation between the size of the pathways and their positions in the ranking; however, not the same can be argued about POR and AOR. The lower performances of PO and AO are due to the harsh punishments for overlapping gene sets; Hence, they select first small pathways which were ranked in the lowest positions by Shapley values alone. We refer to Table~\ref{fig:jaccard_rate} for additional details. \par
\begin{figure}
		\centering
		\includegraphics[width= \textwidth]{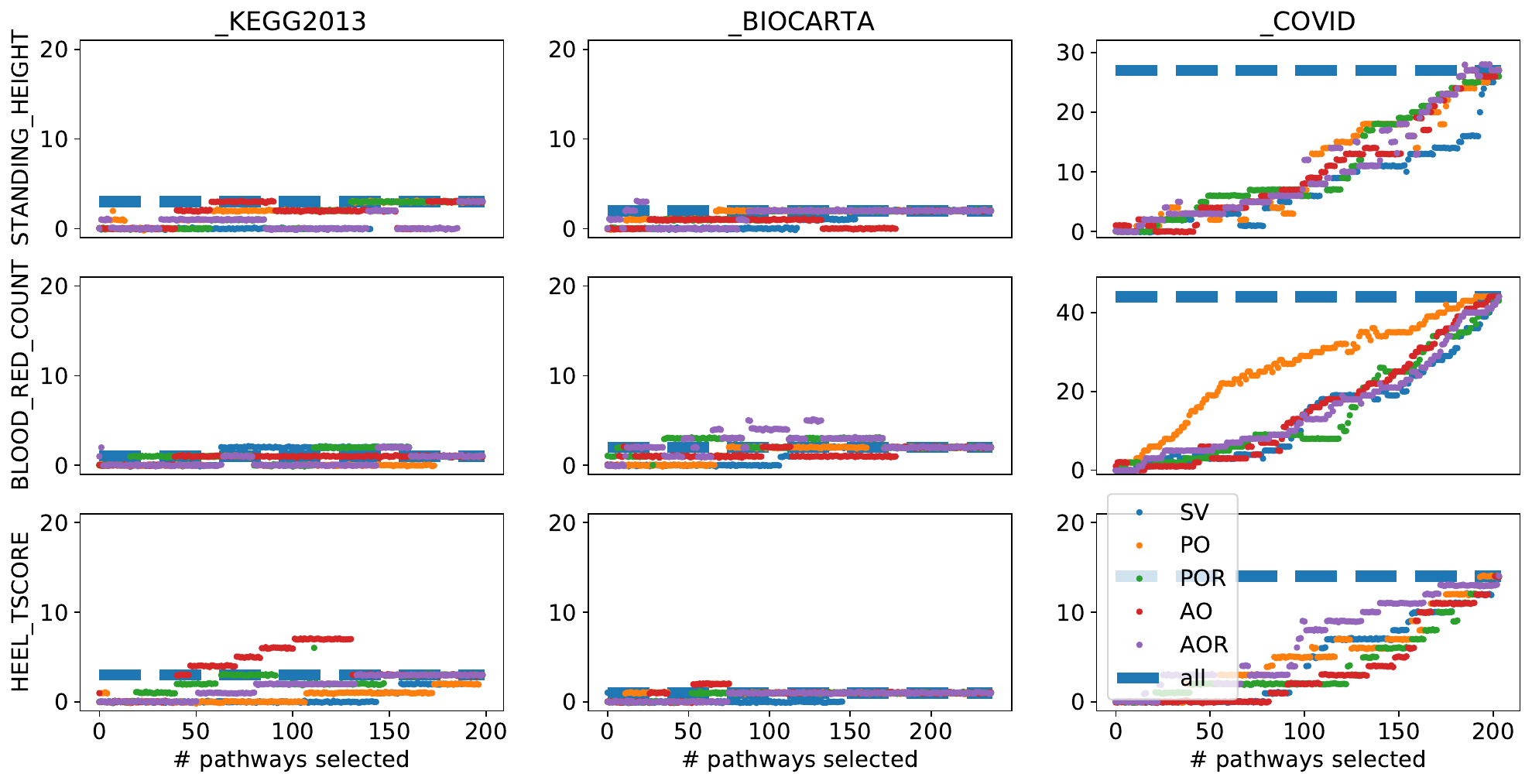}
		\caption{The $y$ axis represents the number of significant pathways for the various proposed rankings (color coded) and for an increasing number of pathways (on the $x$ axis) included in the multiple test corrections.}
		\label{fig:statistical_power_increase}
	\end{figure}

	\textbf{similar statistical power as the whole collections:} We use the Fisher's exact test~\cite{fisher_logic_1935,agresti_introduction_2018} to determine whether a pathway is significant and apply FDR multiple hypothesis testing corrections for the p-values~\cite{benjamini_controlling_1995,dudoit_multiple_2008}. Figure~\ref{fig:statistical_power_increase} shows that the numbers of statistical significant pathways recalled when using only the portions of the collection of gene sets given by the proposed rankings and by the full collections are not remarkably different; we specifically use some remarkable association traits and report the numbers of statistical significant pathways when using the first $40\%$ of the ranked pathways in Table~\ref{tab:enrich_comparison}. For additional experiments with other collections and a comparison with the most common GSEA methods, we refer to Balestra et al.~\cite{balestra2023redundancy}.

	\section{Related work}
	CGT led to applications in computer science in various contexts and applications; In particular, Shapley values have been extended to feature selection~\cite{cohen_feature_2007,pfannschmidt_evaluating_2016,balestraUSV}, networks and security~\cite{van_campen_new_2017}, explainable machine learning~\cite{lundberg_unified_2017} and bioinformatics~\cite{sun_game_2020}. A recent paper by Rozemberczki et al.~\cite{rozemberczki2022shapley} summarizes the major applications of Shapley values in machine learning literature. Moreover, importance scores based on Shapley values have been adapted to study the relationships among genetic and phenotypic characteristics for gene sets prioritization analysis~\cite{lucchetti_shapley_2010,moretti_combining_2008}. Shapley values' exact computation, on the other hand, requires $2^N$ evaluations of a value function where $N$ is the number of players; on the other hand, the introduction of microarray games~\cite{moretti_class_2007} reduces the computational challenges of exact Shapley values' computation to polynomial time under the assumption that the game can be written using only binary relationships. \par
	Collections of gene sets are families of pathways based on some prior biological knowledge~\cite{liberzon_molecular_2015}; the pathways are often overlapping thus making these collections not easily interpretable. Several methods have been proposed to address the low interpretability of the high-dimensional collections, including visualization tools and techniques to merge gene sets into a non-redundant single and unified pathway~\cite{belinky_pathcards_2015,van_iersel_presenting_2008,doderer_pathway_2012}. Stoney et al.~\cite{stoney_using_2018} was the first to point out the importance of maximizing gene coverage without altering the pathways from their original form. However, their work handles redundancy among collections of gene sets in the different databases and \emph{not} in the single collections. \par
	One of the main applications of gene sets is enrichment analyses GSEA~\cite{subramanian_gene_2005,mathur_gene_2018}; GSEA tests for the potential over/under-representation of the analyzed genes in specifically biologically annotated gene sets. The number of statistical hypothesis tests performed equals the number of pathways within the collection of gene sets. Thus, correcting for multiple testing becomes a major challenge~\cite{dudoit_multiple_2008,noble_how_2009}. Among the different corrections for multiple hypothesis testing, we recall the Bonferroni correction and the false discovery rate FDR~\cite{benjamini_controlling_1995,benjamini_control_2001}, where the latter is less conservative.

	% _____________________________________________________________________________ 
	
	\section{Conclusions and discussion}\label{sec:conclusions}
	We proposed redundancy-aware Shapley values to rank sets in families of sets. The four presented rankings aim to satisfy various properties when selecting sets based on them. Motivated by the numerous applications of unsupervised feature selection, the proposed importance scores consider the distribution of elements within the family of sets and their overlap. In the presented application on collections of gene sets, we show good performance for various metrics. However, the range of potential applications is much broader, and our proposed methods can open new research paths in the applied and theoretical fields. 
	For the specific case of collections of gene sets, one possible extension is the addition of a supervised punishment that also considers the relevance of the single pathway to a specific phenotypic trait. This could lead to a higher number of significant pathways when using the Shapley values-based rankings to reduce the dimension of the collections.

	\section{Acknowledgements} 
	{This research was supported by the research training group \emph{Dataninja} funded by the German federal state of North Rhine-Westphalia and by the Research Center Trustworthy Data Science and Security, an institution of the University Alliance Ruhr.}

	% PSEUDOCODE
	
	\bibliographystyle{acm}
	\bibliography{sample-base}

\end{document}